\documentclass[letterpaper]{sig-alternate-10pt}

\usepackage[latin1]{inputenc}
\usepackage[breaklinks,pdfborder={0 0 0}]{hyperref}
\usepackage{algorithm}
\usepackage{algorithmicx}
\usepackage{algpseudocode}
\usepackage{subfigure}
\usepackage{xspace}
\usepackage{ifpdf}

\newcommand{\wrt}{w.r.t.\xspace}
\newcommand{\eg}{e.g.,\xspace}

\newcommand{\ie}{i.e.,\xspace}
\newcommand{\cf}{cf.\xspace}

\newcommand{\MAC}{\textsc{mac}\xspace}

\newcommand{\IEEE}[1]{\mbox{IEEE #1}\xspace}

\newcommand{\Foobar}{Wi\-Pal\xspace}

\newcommand{\myincludegraphics}[2][1]{%
\ifpdf
  \includegraphics[#1]{#2.png}
\else
  \includegraphics[#1]{#2.eps}
\fi
}

\begin{document}

\newcommand{\inserttitle}{%
  \Foobar: Efficient Offline Merging of \IEEE{802.11} Traces}
\hypersetup{pdftitle=\inserttitle, pdfauthor=Thomas Claveirole and
  Marcelo Dias de Amorim}
\title{\inserttitle}

\numberofauthors{2}
\author{
  \alignauthor Thomas Claveirole\\
  \affaddr{LIP6/CNRS~---~UPMC Univ Paris 06}
  \email{\href{mailto:thomas.claveirole@lip6.fr}{thomas.claveirole@lip6.fr}}
  \alignauthor Marcelo Dias de Amorim\\
  \affaddr{LIP6/CNRS~---~UPMC Univ Paris 06}
  \email{\href{mailto:marcelo.amorim@lip6.fr}{marcelo.amorim@lip6.fr}}}

\newcommand{\insertkeywords}{\IEEE{802.11}, Wireless, Trace, Merging,
  Synchronization}
\hypersetup{pdfkeywords=\insertkeywords}

\maketitle

\begin{abstract}
  Merging wireless traces is a fundamental step in mea\-sure\-ment-based
  studies involving multiple packet sniffers. Existing merging tools
  either require a wired infrastructure or are limited in their
  usability. We propose \Foobar, an offline merging tool for
  \IEEE{802.11} traces that has been designed to be efficient and
  simple to use. \Foobar is flexible in the sense that it does not
  require any specific services, neither from monitors (like
  synchronization, access to a wired network, or embedding specific
  software) nor from its software environment (\eg an SQL server).  We
  present \Foobar's operation and show how its features~--- notably,
  its modular design~--- improve both ease of use and
  efficiency. Experiments on real traces show that \Foobar is an order
  of magnitude faster than other tools providing the same features. To
  our knowledge, \Foobar is the only offline trace merger that can be
  used by the research community in a straightforward fashion.
\end{abstract}

\section{Introduction}

Sniffing is a usual technique for monitoring wireless networks. It
consists in spreading within some target area a number of monitors (or
{\it sniffers}) that capture all wireless traffic they hear and
produce traces consisting of \MAC frame exchanges.  Wireless sniffing
is a fundamental step in a number of network operations, including
network diagnosis~\cite{cheng07automating}, security
enhancement~\cite{bahl06enhancing}, and behavioral analysis of
protocols~\cite{cheng06jigsaw, mahajan06analyzing, yeo04framework,
  jardosh05understanding_congestion}.

Wireless sniffing often involves
a centralized process that is responsible for combining the
traces~\cite{cheng06jigsaw, mahajan06analyzing, yeo04framework}. The
objective is to have a global view of the wireless activity from
multiple local measurements. Individual sniffers can also compensate
for their frame losses with data from other sniffers. Merging is
however a difficult task; it requires precise synchronization among
traces (up to a few microseconds) and bearing the unreliable nature of
the medium (frame loss is unavoidable). The literature has provided
the community with a number of merging tool, but they either require
a wired infrastructure or are too specific to the experimentations
conducted in the papers (see more details in
Section~\ref{sec:probem_description})~\cite{cheng06jigsaw,dujovne06wismon,mahajan06wit,mahajan06analyzing}.

In this paper we present \Foobar, an \IEEE{802.11} trace merging tool
that focuses on ease-of-use, flexibility, and speed.  By explaining
\Foobar's design choices and internals, we intend to complete existing
papers and give additional insights about the complex process of trace
merging.  \Foobar has multiple characteristics that distinguish it
from the few other traces mergers:

\begin{description}

\item[Offline tool.] Being an offline tool enables \Foobar to be
  independent of the monitors: one may use any software to acquire
  data. Most trace mergers expect monitors to embed specific
  software~\cite{cheng06jigsaw, dujovne06wismon}.

\item[Independent of infrastructure.] \Foobar's algorithms do not
  expect features from traces that would require monitors to access
  a network infrastructure (\eg synchronization).  Monitors just need
  to re\-cord data in a compatible input format.

\item[Compliant with multiple formats.] \Foobar supports most of the
  existing input formats, whereas other trace mergers require
  a specific format. Some tools even require a custom dedicated
  format~\cite{cheng06jigsaw}.

\item[Hands-on tool.] \Foobar is usable in a straightforward fashion
  by just calling the adequate programs on trace files. Other mergers
  require more complex setups (\eg a database
  server~\cite{mahajan06analyzing} or a network setup involving
  multiple servers~\cite{cheng06jigsaw}.)

\end{description}

This paper provides an analysis that supports these choices (\cf
Section~\ref{sec:eval}).  First, the proposed synchronization
mechanism exhibits better precision than existing algorithms. Second,
\Foobar is an order of magnitude faster than the other publicly
available offline merger, Wit~\cite{mahajan06analyzing}.  This
analysis uses CRAWDAD's uw/sig\-comm\-2004
dataset~\cite{uw-sigcomm2004-2006-10-17}, recorded during the SIGCOMM
2004 conference.\footnote{To the extent of our knowledge, this is the
  only one dataset that is both publicly available and that provides
  enough data to perform merging operations.}  It allows us to
calibrate various parameters of \Foobar, validate its operation, and
show its efficiency. \Foobar is however \emph{not} designed for
a specific dataset and works on any wireless traces using the
appropriate input format (\Foobar's test suite includes various
synthetic traces with different formats).  We do believe that \Foobar
will be of great utility for the research community working on
wireless network measurements.

\section{Trace merging: overview}
\label{sec:probem_description}

Wireless sniffing requires the use of multiple monitors for {\it
  coverage} and {\it redundancy} reasons. Coverage is concerned when
the distance between the monitor and at least one of the transmitters
to be sniffed is too large to ensure a minimum reception
threshold. Redundancy is the consequence of the unreliability of the
wireless medium. Even in good radio conditions monitors may miss
successfully transmitted frames. After the collection phase, traces
must be combined into one. A merged trace holds all the frames
recorded by the different monitors and gives a global view of the
network traffic.

\begin{figure}[t]
  \centering
  \fbox{\begin{minipage}{.95\linewidth}
      \centering
      \medskip
      \myincludegraphics[width=.7\linewidth]{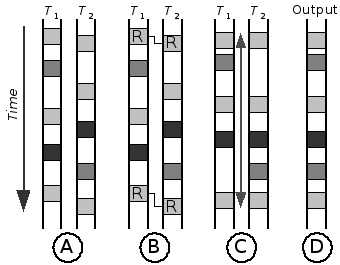}
      \footnotesize
      \begin{enumerate}
      \item[A.] The traces are not synchronized and miss some frames.
      \item[B.] One identifies some reference frames common to both
        traces. This information enables trace synchronization.
      \item[C.] One adjusts the frames' timestamps and synchronize $T_1$
        and $T_2$.
      \item[D.] One can merge the traces. Duplicate frames are only
        accounted once.
      \end{enumerate}
      \smallskip
    \end{minipage}}
  \caption{Merging two traces $T_1$ and $T_2$.}
  \label{fig:basics}
\end{figure}

The traditional approach to merging traces involves a {\it
  synchronization} step, which aligns frames according to their
timestamps. This enables identifying all frames that are identical in
traces so that they appear once and only once in the output trace
(Cheng et al~\cite{cheng06jigsaw} refer to it as {\it unification}.)
This process is illustrated in Fig.~\ref{fig:basics}.

Synchronization is difficult to obtain because, in order to be useful,
it must be very precise. Imprecise frame timestamps may result in
duplicate frames and incorrect ordering in the output trace. An
invalid synchronization may also lead to distinct frames accounted for
the same frame in the output trace. In order to avoid such undesirable
effects one needs precision of less than
$106\mu{}s$~\cite{yeo04framework}. To the extent of our knowledge, no
existing hardware supports synchronizing network cards' clocks with
such a precision (note that we are interested in frame arrival times
in the card, not in the operating system).

Therefore, all merging tools post-process traces to {\it
  resynchronize} them with the help of {\it reference frames}, which
are frames that appear in multiple traces. One may readjust the
traces' timing information using the timestamps of the reference
frames (see Fig.~\ref{fig:basics}.) Finding reference frames is
however a hard task, since we must be sure a given reference frame is
an occurrence of the same frame in every traces. That is, some frames
that occur frequently (\eg \MAC acknowledgements) cannot be used as
reference frames because their content does not vary
enough. Therefore, only a subset of frames are used as reference
frames, as explained later in this paper (\cf
Section~\ref{sec:details}).

A few trace merging tools exist in the literature, but they do not
focus on the same set of features as this paper.  For instance,
Jigsaw~\cite{cheng06jigsaw} is able to merge traces from hundreds of
monitors, but requires monitors to access a network
infrastructure. WisMon~\cite{dujovne06wismon} is an online tool that
has similar requirements.  This paper however considers smaller-scale
systems (dozens of monitors) but where no monitor can access a network
infrastructure.  Another system close to ours is
Wit~\cite{mahajan06wit, mahajan06analyzing}.  Despite Wit provides
valuable insights on how to develop a merging tool, it is difficult to
use, modify, and extend in practice (\cf authors' note in
CRAWDAD~\cite{mahajan06wit}).  Thus our motivation to propose a new
trace merger.  Note that this paper only refers to Wit's merging
process (as Wit has other features like, \eg a module to infer missing
packets).

\section{\Foobar's basics}
\label{sec:basics}

\Foobar has been designed according to the following constraints:

\begin{description}

\item[No wired connectivity.] The sniffers must be able to work in
  environments where no wired connectivity is provided. This enables
  performing measurements when it is difficult to have all sniffers
  access a shared network infrastructure (\eg in some conference
  venues, or when studying interferences between two wireless networks
  belonging to distinct entities).

\item[Simplicity to the end-user.] We believe simplicity is the key to
  re-usability. Users are not expected to install and set up complex
  systems (\eg a da\-ta\-base backend) in order to use \Foobar.

\item[Clean design.] \Foobar exhibits a modular design. Developers can
  easily adapt part of the trace merger (\eg the reference frames
  identification process, the synchronization, or merging algorithm.)

\end{description}

For these reasons, we opted for an offline trace merger that does not
require that traces be synchronized a priori. Concretely, the sniffers
only have to record their measurements on a local storage device,
using the wide\-ly used PCAP (Packet CAPture) file format. \Foobar
comes as a set of binaries to manipulate wireless traces, including
the merging tool presented in this paper. It works directly on PCAP
files both as input and output. \Foobar is composed of roughly 10k
lines of C++ and makes heavy usage of modern generic and static
programming techniques. \Foobar is downloadable from
\url{http://wipal.lip6.fr}.

\begin{figure}[t]
 \centering
 \myincludegraphics[width=7cm]{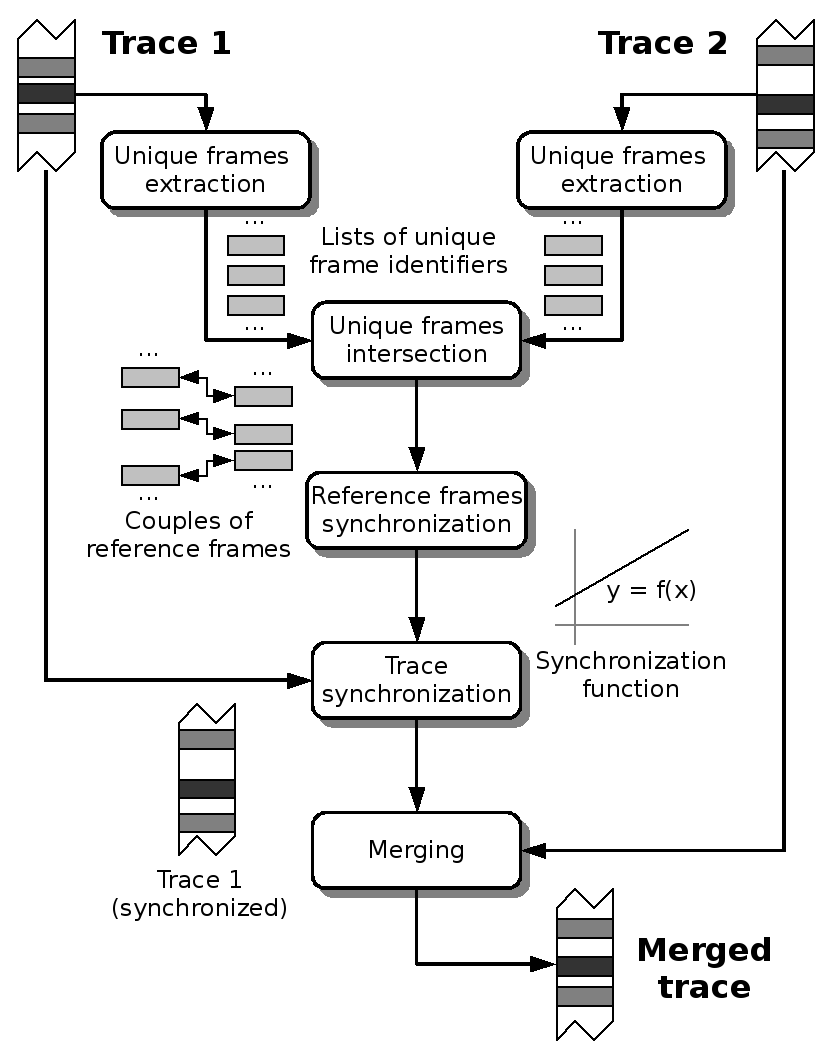}
 \caption{\Foobar's overall structure.}
 \label{fig:design}
\end{figure}

\section{\Foobar's detailed operation}
\label{sec:details}

Fig.~\ref{fig:design} depicts \Foobar's structure.  Each box
represents a distinct module and arrows show \Foobar's data
flow. \Foobar takes two wireless traces as input and produces a single
merged trace.\footnote{In order to merge more than two traces, it
  suffices to execute the merging tool as many times as required (two
  by two).} In the following, we explain in detail the functioning of
each one of the modules.

\subsection{Identifying reference frames}
\label{sec:unique-frames}

This section explains the process of extracting reference frames. This
operation involves two steps: extraction of unique frames and
intersection of unique frames (see Fig.~\ref{fig:design}.)

Let us first define what a unique frame means. A frame is said to be
unique when it appears ``in the air'' once and only once for the whole
duration of the measurement. A frame that is unique within each trace
but that actually appeared twice on the wireless medium should not be
considered as unique.

The process of extracting unique frames finds candidates to become
reference frames. The process of intersecting unique frames identifies
then identical unique frames from both traces to become reference
frames.

\subsection{Extraction of unique frames}
\label{sec:unique-frames:extract}

\Foobar consider every beacon frame and non-retrans\-mit\-ted probe
response as a unique frame. These are management frames that access
points send on a regular basis (\eg every 100 ms for beacon
frames). The uniqueness of these frames is due to the 64-bit
timestamps they embed (these timestamps are not related to the actual
timestamps used for synchronization).

In practice, the extraction process does not load full frames into
memory. It uses 16-byte hashes instead, which are stored in memory and
used for comparisons. Limiting the size of stored information is an
important aspect since, as we will see later, \Foobar's intersection
process performs a lot of comparisons and needs to store many unique
frames in memory. Tests with CRAWDAD's uw/sigcomm2004
dataset~\cite{uw-sigcomm2004-2006-10-17} have shown that this
technique is practical. Concretely, \Foobar needs less than 600~MB to
load 7,700,000 unique frames.

There are some rare cases where the assumption that beacons and probe
responses are unique does not hold. The uw/sigcomm2004 dataset has
a total number of 50,375,921 unique frames (about 14\% of 364,081,644
frames). Among those frames, we detected 5 collisions (distinct unique
frames sharing identical hashes.) \Foobar's intersection process
includes a filtering mechanism to detect and filter such collisions
out.

\subsection{Intersection}
\label{sec:unique-frames:intersect}

The intersection process intersects the sets of unique frames from
both input traces. There are multiple algorithms to perform such
a task. Based on Cheng et al.~\cite{cheng06jigsaw}, a solution is to
``bootstrap'' the system by finding the first unique frame common to
both traces and then use this reference frame as a basis for the
synchronization mechanism, as shown in
Algorithm~\ref{alg:inter-sync}. One may also use subsequent reference
frames to update synchronization. This algorithm is practical because
the inner loop only searches a very limited subset of $I_2$. It has
several drawbacks though: (i) the performance of the algorithm
strongly depends on the precision of the synchronization process; (ii)
finding the first reference frame is still an issue; (iii) this
algorithm couples intersection with synchronization, which is
undesirable with respect to modularity; and (iv) there is
a possibility that some frames are read multiple times from
$I_2$. More specifically, access to $I_2$ is not sequential.

\begin{algorithm}[t]
 \footnotesize
 \begin{algorithmic}
 \Statex \textbf{Input:} two lists of unique frames $I_1$ and $I_2$.
 \Statex \textbf{Output:} a list of reference frames.
 \Statex
 \State $\delta \leftarrow$ synchronization precision
 \ForAll{$u_1 \in I_1$}
 \State $t_{u_1} \leftarrow u_1\text{'s time of arrival}$
 \ForAll{$u_2 \in I_2$ between $t_{u_1} - \delta$ and $t_{u_1} + \delta$}
 \If{$u_2$ is an occurrence of $u_1$}
 \State Append $(u_1, u_2)$ to output.
 \EndIf
 \EndFor
 \EndFor
 \end{algorithmic}
 \caption{Intersection using synchronization.}
 \label{alg:inter-sync}
\end{algorithm}

\Foobar includes an algorithm that is much simpler to implement and
that avoids the drawbacks of the abovementioned solution. The main
characteristics of the proposed algorithm (detailed in
Algorithm~\ref{alg:inter-nosync}) are: (i) it does not require
a bootstrapping phase; (ii) it does not depend on any kind of
synchronization; and (iii) It sequentially reads each frame only once
from $I_1$ and $I_2$.

The algorithm starts by loading all unique frames of the first trace
into memory. This precludes using it as an online tool. Note that
loading all unique frames from a trace into memory may hog resources;
this justifies the importance of having small identifiers for the
unique frames. These constraints are however negligible compared to
those of Algorithm~\ref{alg:inter-sync}. To support our argument, let
us show an example using the uw/sig\-comm\-2004 dataset. The biggest
traces are those from sniffers {\it mojave} and {\it sonoran} on
channel 11 (roughly 19~GB each.) Extracting these traces' unique
frames and intersecting them using \Foobar needs 575 MB of
memory. Therefore, memory aggressiveness is not a concern in
Algorithm~\ref{alg:inter-nosync}.

\begin{algorithm}[t]
 \footnotesize
 \begin{algorithmic}
 \Statex \textbf{Input:} two lists of unique frames $I_1$ and $I_2$.
 \Statex \textbf{Output:} a list of reference frames.
 \Statex
 \State $h \leftarrow \emptyset$ \Comment{Implement $h$ with a hash table.}
 \ForAll{$u_1 \in I_1$}
 \State Insert $u_1$ into $h$.
 \EndFor
 \ForAll{$u_2 \in I_2$}
 \If{$h$ contains an occurrence $u_1$ of $u_2$}
 \State Append $(u_1, u_2)$ to output.
 \EndIf
 \EndFor
 \end{algorithmic}
 \caption{\Foobar's intersection algorithm.}
 \label{alg:inter-nosync}
\end{algorithm}

Another advantage of Algorithm~\ref{alg:inter-nosync} is its ability
to detect collisions of unique frames within the first
trace. Collisions are detected by duplicate elements in $h$. \Foobar
detects such cases, memorizes collisions, and filter them out of the
hash table before starting the algorithm's second loop. Of course,
collisions in the second trace remain undetected. Even if \Foobar
detected them, there would still be the possibility that a collision
spans across both traces (\ie each trace contains one occurrence of
a colliding unique frame). Such cases lead to producing invalid
reference frames. To detect them, \Foobar looks at possible anomalies
\wrt the interarrival times between unique frames. In practice,
invalid references are rare: only three occurrences when merging
uw/sig\-comm\-2004's channel 11 (a 73~GB input which produces a 22~GB
output).

\subsection{Synchronization}
\label{sec:synchronize-merge:sync}

Synchronizing two traces means mapping trace one's timestamps to
values compatible with trace two's.  \Foobar computes such a mapping
with an affine function \mbox{$t_2 = a\:t_1 + b$}.  It estimates $a$
and $b$ with the help of reference frames as the process runs.

\Foobar's synchronization process operates on windows of $w + 1$
reference frames (finding an optimal value of $w$ is discussed
below). For each reference frame $R_i$, the process performs a linear
regression using reference frames $R_{i-\lfloor w/2 \rfloor}$, \ldots,
$R_{i+\lceil w/2 \rceil}$. At the beginning and at the end of the
trace, we use \mbox{$R_1, \ldots, R_w$} and \mbox{$R_{N - w}, \ldots,
  R_N$} ($N$ is the number of reference frames.)  The result gives $a$
and $b$ for all frames between $R_i$ and $R_{i+1}$.

\begin{figure}
 \centering
 \myincludegraphics[width=8cm]{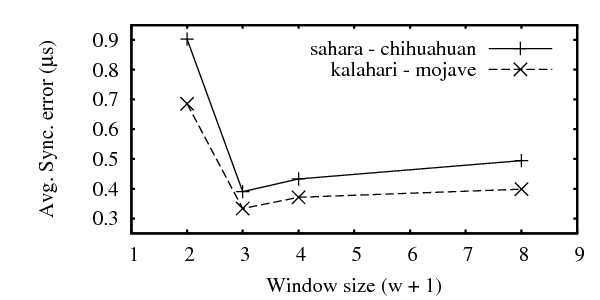}
 \caption{Average synchronization error \wrt linear regression window size.}
 \label{fig:prec}
\end{figure}

We performed a number of experiments that revealed that the optimal
value for $w$ is 2 (\ie \Foobar performs linear regressions on 3-frame
windows). Fig.~\ref{fig:prec} shows the results of performing two
merge operations with varying window sizes. The merges concern channel
11 of the {\it sahara~-- chihuahuan} and {\it kalahari~-- mojave}
sniffers from uw/sig\-comm\-2004. The average synchronization error is
computed as follows. Consider only the subset $S$ of frames that are
shared by both the first and second trace $T_1$ and $T_2$. For a given
frame $f$, let $t_{f, 1}$ be the arrival time of $f$ inside $T_1$
(after clock synchronization) and $t_{f, 2}$ be the arrival time of
$f$ inside $T_2$. The average synchronization error is given by
\mbox{$\frac{1}{|S|}\sum_{f \in S}|t_{f, 2} - t_{f, 1}|$}. As
previously underlined, $w = 2$ leads to the minimum average
synchronization error. Note that techniques that use $w = 1$ (\ie that
performs linear interpolations on couples of reference frames) would
lead to the worst synchronization error. Furthermore, merging traces
with $w = 1$ misses some shared frames. Table~\ref{tab:sync} shows the
number of frames that are identified as duplicates in the input
traces. Whereas using $w > 1$ always gives identical results, using $w
= 1$ leads to some missed duplicates (7,455 for {\it sahara~--
  chihuahuan} and 84 for {\it kalahari~-- mojave}). Although this is
a small number compared to the total number of frames in the output
traces, it indicates that synchronizing traces using linear
interpolation (as Wit~\cite{mahajan06wit} does) may lead to incorrect
results. Unfortunately, it is difficult to know whether some
duplicates were missed when $w > 1$ (we do not know which frames to
expect as duplicates).

\begin{table}[t]
 \footnotesize
 \centering
 \begin{tabular}{|c|c|c|}
 \hline
 & \multicolumn{2}{c|}{Number of shared frames} \\
 & \multicolumn{1}{c}{$w = 1$} & \multicolumn{1}{c|}{$w > 1$} \\
 \hline\hline
 {\it sahara} -- {\it chihuahuan} & 32,312,812 & 32,320,267 \\
 {\it kalahari} -- {\it mojave} & 840,143 & 840,227 \\
 \hline
 \end{tabular}
 \caption{Number of frames found to be shared by both input traces when
   merging {\it sahara~-- chihuahuan} and {\it kalahari~-- mojave} with
   $w = 1$ and $w > 1$ (channel 11).}
 \label{tab:sync}
\end{table}

\subsection{Merging}
\label{sec:synchronize-merge:merge}

We now present how \Foobar performs the final step, namely the merging
process itself. Its role is to copy frames from synchronized traces to
the output trace. Of course, it must order its output correctly while
avoiding duplicate frames.

Algorithm~\ref{alg:merge} details \Foobar's merging algorithm. For the
sake of illustration, we present here a simplified version that
assumes that only one frame is emitted at a given time inside the
monitoring area. It simultaneously iterates on both inputs, where each
iteration adds the earliest input frame to the output
(lines~\ref{alg:merge:it-1} and~\ref{alg:merge:it-2}.) Duplicate
frames are the ones that have identical contents and that are spaced
less than 106$\mu{}s$ (line~\ref{alg:merge:dup}.) The rationale for
this value is that 106$\mu{}s$ is half of the minimum gap between two
valid \IEEE{802.11} frames~\cite{yeo04framework}. Therefore, the
appearance of identical frames during such an interval is in fact
a unique occurrence of the same frame.

\begin{algorithm}[t]
 \footnotesize
 \begin{algorithmic}[1]
 \Statex \textbf{Input:} two synchronized traces $T_1$ and $T_2$.
 \Statex \textbf{Output:} the merge of $T_1$ and $T_2$.
 \Statex
 \Procedure{Advance}{$f$: frame, $T$: trace}
 \State Append $f$ to output; $f \leftarrow T$'s next frame (or $nil$)
 \EndProcedure
 \Statex
 \State $f_1 \leftarrow T_1$'s first frame; $f_2 \leftarrow T_2$'s first
 frame
 \While{$f_1 \neq nil$ \textbf{or} $f_2 \neq nil$}
 \If{$f_1 = nil$} \textsc{Advance}($f_2$, $T_2$)
 \ElsIf{$f_2 = nil$} \textsc{Advance}($f_1$, $T_1$)
 \Else
 \State $t_{f_1} \leftarrow f_1$'s time of arrival
 \State $t_{f_2} \leftarrow f_2$'s time of arrival
 \If{$f_1$ = $f_2$ \textbf{and} $|t_{f_1} - t_{f_2}| < 106\mu{}s$}
 \label{alg:merge:dup}
 \State Append either $f_1$ or $f_2$ to output.
 \State $f_1 \leftarrow T_1$'s next frame (or $nil$)
 \State $f_2 \leftarrow T_2$'s next frame (or $nil$)
 \ElsIf{$t_{f_1} < t_{f_2}$} \textsc{Advance}($f_1$, $T_1$)
 \label{alg:merge:it-1}
 \Else{} \textsc{Advance}($f_2$, $T_2$)
 \label{alg:merge:it-2}
 \EndIf
 \EndIf
 \EndWhile
 \end{algorithmic}
 \caption{\Foobar's merging algorithm.}
 \label{alg:merge}
\end{algorithm}

\section{Evaluation}
\label{sec:eval}

This section provides an evaluation of \Foobar using CRAWDAD's
uw/sigcomm2004 dataset~\cite{uw-sigcomm2004-2006-10-17}. We
investigate both the correctness and the efficiency of \Foobar. We
merge all traces sniffed from channel 11 and then use some heuristics
to evaluate the quality of the result. We also analyze \Foobar's
speed.

\begin{figure}
 \centering
 \myincludegraphics[width=8cm]{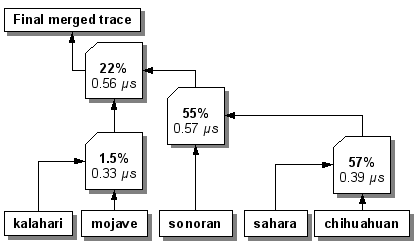}
 \caption{Summary of uw/sigcomm2004's merging process, channel 11.
   Percentages indicate the number of frames that are shared by parent
   traces. Bottom figures indicate the average synchronization error.}
 \label{fig:eval:merge}
\end{figure}

Traces from five sniffers compose the uw/sig\-comm\-2004 dataset: {\it
  chihuahuan}, {\it kalahari}, {\it mojave}, {\it sahara}, and {\it
  so\-no\-ran}. Fig.~\ref{fig:eval:merge} shows the merging sequence
we used to merge all traces. The reason why {\it kalahari} and {\it
  mojave} share so few frames is that {\it kalahari} is an order of
magnitude smaller than {\it mojave}.

\subsection{Correctness}
\label{sec:correctness}

Checking the correctness of the output is difficult. Being able to
test whether traces are correctly merged or not would be equivalent to
knowing exactly in advance what the merge should look
like. Unfortunately, there is no reference output against which we
could compare. Thus, we propose several heuristics to check if \Foobar
introduces or not inconsistencies in its outputs. We also check
\Foobar's correctness with a test-suite of synthetic traces for which
we know exactly what to expect as output.

A broken merging process could lead to several inconsistencies in the
output traces.  Regarding the uw/sig\-comm\-2004 dataset, we
investigate in particular two of those inconsistencies: duplicate
unique frames and duplicate data frames.

\begin{description}

\item[Duplicate unique frames.] As seen previously, every unique frame
  should only occur once in the traces (including merged traces). Yet,
  it is difficult to avoid collisions in practice (see
  Section~\ref{sec:unique-frames:extract}). Thus one should not
  consider all collisions as inconsistencies. When merging
  uw/sigcomm2004, the final trace has 5 collisions. We manually
  verified that they are not inconsistencies introduced by \Foobar's
  merging process.

\item[Duplicate data frames.] We search traces on a per-sender basis
  for successive duplicate data frames (only considering
  non-retransmitted frames). Such cases should not occur in theory~--
  without retransmissions sequence numbers should at least
  vary. Surprisingly, traces from uw/sigcomm2004 contain 20,303 such
  anomalies. We have no explanations why the dataset exhibits those
  phenomena. We checked however that the merged trace does not have
  more duplicates than the original traces.

\end{description}

\subsection{Efficiency}
\label{sec:efficiency}

Merging all the traces (73~GB) takes about 2 hours and 20 minutes
(real time) on a 3~GHz processor with 2~GB RAM. We balance merge
operations on two hard drives, whose average throughput during
computations are about 60~MB/s and 30~MB/s. The average CPU usage is
75\%, which means one could perform faster with faster hard drives
(about 1 hour and 40 minutes).

Comparing \Foobar with online trace mergers does not make much sense:
their mode of operation is different, and these also have different
requirements (\eg wired connectivity and loose synchronization.) The
comparison would be unfair. We can however compare \Foobar with
Wit~\cite{mahajan06wit}, another offline merger. Wit works on top of
a database backend, which means that trace files need to be imported
into a database before any further operation can begin (\eg merging or
inferring missing packets). Using the same machine as before,
importing channel 11 of uw/sigcomm2004 into Wit's database takes
around 33 hours (user time). This means that, before Wit begins its
merge operations, \Foobar can perform at least 14 runs of a full merge
with the same data. \Foobar allows then tremendous speed
improvements. One of the reasons for such a difference is \Foobar uses
high performance C++ code while Wit is just a set of Perl scripts
using SQL to interact with a database.

\section{Conclusion}
\label{sec:discussion}

This paper introduced the \Foobar trace merger. As an {\it offline}
merger, \Foobar does not require sniffers to be synchronized nor to
have access to a wired infrastructure. \Foobar provides several
improvements over existing equivalent software: (i) it comes as
a simple program able to manipulate trace files directly, instead of
requiring a more complex software setup, (ii) its synchronization
algorithm offer better precision than the existing algorithms; and
(iii) it has a clean modular design. Furthermore, we also showed
\Foobar is an order of magnitude faster than Wit~\cite{mahajan06wit},
the other available offline merger.

We have several plans for the future of \Foobar. First, we are
currently extending it to include other features (besides merging). As
a flavor of future features of \Foobar, it will perform traffic
statistics on \IEEE{802.11} traces. We will also make better use of
\Foobar's modularity and test other algorithms for the various stages
of the merging operation.

\end{document}